%% file: main.tex
\documentclass[11pt]{article}

\usepackage[preprint]{acl}

\usepackage{times}
\usepackage{latexsym}

\usepackage[T1]{fontenc}

\usepackage[utf8]{inputenc}

\usepackage{microtype}

\usepackage{inconsolata}

\usepackage{graphicx}

\usepackage{todonotes}

\usepackage{booktabs} 

\usepackage{amsmath}
\usepackage{amssymb}
\usepackage{mathtools}
\usepackage{amsthm}

\usepackage{algorithmic}            
\usepackage{algorithm}

\usepackage{listings}
\usepackage{xcolor}

\definecolor{codegreen}{rgb}{0,0.6,0}
\definecolor{codegray}{rgb}{0.5,0.5,0.5}
\definecolor{codepurple}{rgb}{0.58,0,0.82}
\definecolor{backcolour}{rgb}{0.95,0.95,0.92}

\lstdefinestyle{mystyle}{
    backgroundcolor=\color{backcolour},   
    commentstyle=\color{codegreen},
    keywordstyle=\color{magenta},
    numberstyle=\tiny\color{codegray},
    stringstyle=\color{codepurple},
    basicstyle=\ttfamily\footnotesize,
    breakatwhitespace=false,         
    breaklines=true,                 
    captionpos=b,                    
    keepspaces=true,                 
    numbers=left,                    
    numbersep=5pt,                  
    showspaces=false,                
    showstringspaces=false,
    showtabs=false,                  
    tabsize=2,
    frame=single,
    framesep=5pt,
    framerule=0.5pt,
    xleftmargin=0.05\columnwidth,     
    xrightmargin=0.05\columnwidth,
    breaklines=true,                 
    breakatwhitespace=false,         
    columns=fullflexible,            
    basicstyle=\ttfamily\footnotesize,   
}

\title{Multi-task LLMs for Bug Classification: \\Efficient Inference with Auxiliary Decoding Heads}

\author{
 \textbf{Nikolai Rozanov\textsuperscript{1,}\footnotemark
 }
\\
\\
 \textsuperscript{1}Department of Computing, Imperial College London, London, UK
\\
 \small{
   \textbf{Correspondence:} \href{mailto:nikolai.rozanov13@imperial.ac.uk}{nikolai.rozanov13@imperial.ac.uk}
 }
}

\begin{document}
\maketitle

\input{sections/0_abstract.tex}

\input{sections/1_introduction}
\input{sections/2_background}
\input{sections/3_method}
\input{sections/4_experiments}

\input{sections/5_results}

\input{sections/6_analysis}

\input{sections/7_conclusion}

\input{sections/8_other}

\bibliography{bibliography}

\appendix
\input{sections/a_appendix}

\end{document}

%% file: sections/0_abstract.tex
\begin{abstract}
The rapid adoption of LLM-powered code generation has dramatically accelerated software development, yet effective verification methods remain severely underdeveloped. Existing bug localization techniques are either prohibitively expensive, requiring minutes of agentic reasoning and thousands of generated tokens per file, and/or operate at coarse function-level granularity unsuitable for precise debugging. While works that focus on line-level granularity and are more light-weight are often limited in their performance or context size. We introduce a novel line-level bug localization approach that addresses these limitations through three key contributions: (1) a token alignment algorithm that overcomes fundamental tokenization challenges in previous work, (2) a lightweight multi-task LLM for bug localization (MLC) enabling efficient line-level bug classification, and (3) an optimized training recipe for multi-line prediction. Our method achieves state-of-the-art performance among similar setups on line-level bug localization with full-file context. At the same time we reach comparable performance to agentic approaches on Defects4J and PypiBugs benchmarks while reducing inference latency by orders of magnitudes, requiring only a single generated token per file. We further demonstrate strong generalization by introducing and evaluating on a small out-of-domain evaluation datasets in Python. We will open source our code, models, and datasets upon acceptance.
\end{abstract}


%% file: sections/1_introduction.tex

\section{Introduction}
\renewcommand{\thefootnote}{\fnsymbol{footnote}}
\footnotetext[1]{This work was partially accomplished at Recurse Ltd., London UK.}
\renewcommand{\thefootnote}{\arabic{footnote}}
With the advancement of LLMs, code generation went beyond just the human software engineers \cite{chen2021evaluatinglargelanguagemodels}. In recent years there was an explosion of autonomous coding systems (i.e. `vibe' coding platforms, LLM Coding Agents, etc.) \cite{geminiteam2025geminifamilyhighlycapable, xia2024agentlessdemystifyingllmbasedsoftware, wang2025openhandsopenplatformai, deepseekai2025deepseekr1incentivizingreasoningcapability}. However, the ability to verify this code lags behind and was not significantly extended beyond initial ML based fault localisation, human reviewers and existing analysis tools \cite{charoenwet2024empiricalstudystaticanalysis}. This creates an enormous problem for both the safety and security of open source software \cite{kharma2025securityqualityllmgeneratedcode} as well as the actual performance and stability of software \cite{ouyang2025kernelbenchllmswriteefficient}.

In our work we are investigating how to address the issue of automatic bug localization. Historically, fault localization either was done using expensive human review, automatic dynamic analysis tools \cite{FaultLocalizationSurvey7390282}, which requires running the code and can be very expensive and resource intensive; or static analysis \cite{Mehrpour2022, FaultLocalizationSurvey7390282, charoenwet2024empiricalstudystaticanalysis}, which can be limited in what can be extracted. To address these limitations LLM based analysis tools were proposed. Our work explores LLM based bug localization.

Existing work focuses primarily on agentic systems, reasoning models, RAG-based systems or classical supervised fine-tuning \cite{chen2025locagentgraphguidedllmagents, asad2026explorativeirblcombiningsemantic,ji2024impactlargelanguagemodels, yang2023largelanguagemodelstestfree, jambigi2025faultlocalizationfinetuninglarge, Qin2024AgentFLSL}. We observe two main issues with existing systems and work, firstly it is very expensive to run such systems \footnote{Popular bug identification online tools like code rabbit, claude caude, recurse etc.} the classification of a single file takes 1 to 2 minutes of LLM inference (and many API calls) and secondly these often work on the function level granularity, which can be limiting for downstream tasks. Additionally, works \cite{yang2023largelanguagemodelstestfree, ji2024impactlargelanguagemodels} that attempted line-level prediction use an additional transformer architecture on top of a standard LLM making it hard and expensive to train or used short context windows at the method level. 




\begin{figure}[t]
\centering
\includegraphics[width=0.9\linewidth]{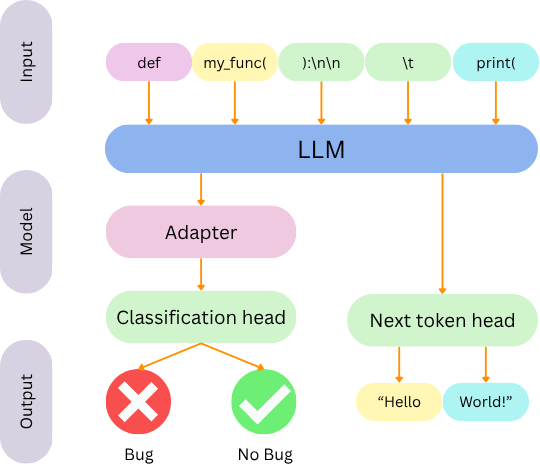}
\caption{Model architecture showing both the original LLM `channel' for next token prediction as well as the additional classification head.}
\label{fig:multi_channel_architecture}
\end{figure}

\begin{figure*}[ht!]
\centering
\includegraphics[width=0.9\linewidth]{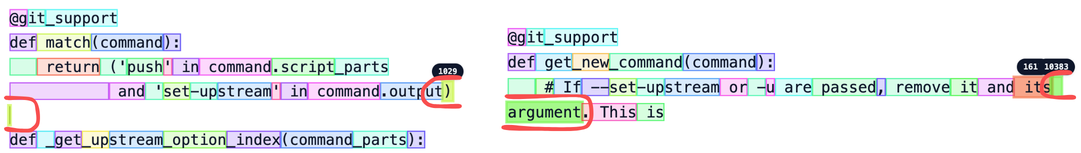}
\caption{Code snippet showing how GPT-5 tokenizer splits the token. Particularly, one can see that the newline `\textbackslash n' token is split over multiple lines and includes other words from either line.}
\label{fig:new_line_tokens}
\end{figure*}

We take a different approach and introduce `Multi-task LLM for Bug Localization' (MLC). Our method achieves the state-of-the-art results on a comparable setup (line-level granularity and full file context) with models ranging from 1.7B to 8B parameters. On the `shorter context' setup we beat the previous state-of-the-art LLMAO and achieve comparable results to Agentic setups. At the same time our method only requires a single token to be generated for all lines to be classified; therefore making inference very fast (especially relative to reasoning and agentic systems that often require multiple thousand tokens to be generated).

Additionally, we introduce a small evaluation dataset of unseen python bugs to evaluate out-of-distribution performance. We observe that general bug localization abilities generalize well, while catching all bugs becomes harder in the OOD domain.

In summary, our contributions are: 1) a Token Alignment Algorithm that overcomes theoretical assumptions made by previous work, 2) a Multi-task LLM modelling technique for bug localization, 3) a training recipe that produces very competitive results at a fraction of the cost and 4) an additional evaluation dataset for out-of-distribution testing.

%% file: sections/2_background.tex
\section{Background and related Work}
\label{sec:background}

Automated Fault Localization (FL) is a core challenge in software debugging, aiming to identify the specific code responsible for a failure. Apart from humans analysing vast amounts of textual data, traditional techniques rely primarily on dynamic analysis, such as Spectrum-Based Fault Localization (SBFL) \cite{FaultLocalizationSurvey7390282}, which depends on expensive test execution to gather coverage spectra; or static analysis like ochiai and others \cite{Mehrpour2022, charoenwet2024empiricalstudystaticanalysis}, which can be limited in scope. The advent of Large Language Models (LLMs) has opened a new paradigm for learning-based fault localization, though existing approaches struggle with efficiency, cost, and granularity.

\textbf{Agentic and retrieval-based localization:} A significant trend frames FL as an information retrieval (IRBL) or agentic reasoning task. Works like \textsc{GenLoc} \cite{asad2026explorativeirblcombiningsemantic}, \textsc{MemFL} \cite{Yeo2025ImprovingLF}, and \textsc{OrcaLoca} \cite{Yu2025OrcaLocaAL} employ multi-step agents, external memory, or graph-based code representations to retrieve relevant files or functions. Similarly, \textsc{AgentFL} \cite{Qin2024AgentFLSL} uses a multi-agent pipeline for comprehension, navigation, and confirmation. While these systems can navigate large codebases, their agentic nature makes them prohibitively expensive for widespread use, often requiring minutes of reasoning, multiple LLM calls, and generating thousands of tokens per bug. Furthermore, their file or function-level granularity is insufficient for automated repair.

\textbf{Line-level prediction and its costs:} For practical debugging, line-level precision is essential. \textsc{LLMAO} \cite{yang2023largelanguagemodelstestfree} directly addresses this by fine-tuning a separate neural network on top of a frozen LLM backbone to predict faulty lines, setting strong Top-k benchmarks. However, this architecture requires training and maintaining an additional transformer model, increasing complexity and cost. It also typically operates within a limited context window (e.g., a single function), missing broader file-level context. Other approaches, such as \textsc{FlexFL} \cite{xu2025flexflflexibleeffectivefault} or studies on fine-tuning LLMs for FL \cite{ji2024impactlargelanguagemodels, jambigi2025faultlocalizationfinetuninglarge}, often rely on generating natural language tokens (CoT), thereby inheriting the large latency.

\textbf{Our approach, efficient direct classification:} Our work diverges from these paradigms. We avoid the high latency of agentic systems and the architectural overhead of attached networks. Instead, we introduce a lightweight fine-tuning method that 
adds and optional adapter and a decoding head to the the LLM's final layer to perform direct, token-level binary classification (\texttt{Bug/No-Bug}) across all lines in a file simultaneously. This enables a single forward pass, requiring only one generated token for the entire file. This significantly reduces cost and latency compared to iterative agents or generative pipelines. Our method achieves precise line-level localization while being more parameter-efficient, easier to train, and capable of leveraging full-file context, addressing key limitations of prior methods. Crucially, the original capability of the LLM remains intact and inference using the original classification head has a negligible overhead.

%% file: sections/3_method.tex
\section{Method}
\begin{figure*}[t!]
\centering
\includegraphics[width=0.9\linewidth]{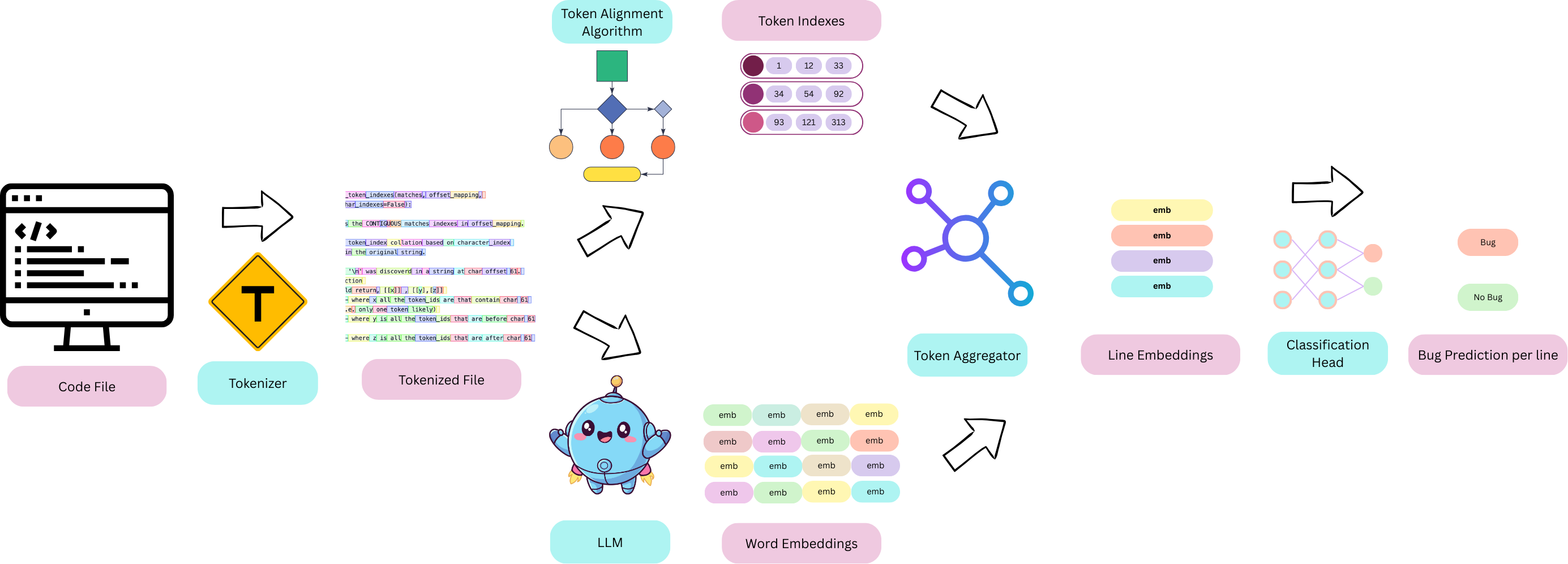}
\caption{The full flow of a code file to a per-line bug classification.}
\label{fig:classification_flow}
\end{figure*}

Overall, our method consists of i) a token alignment algorithm, ii) an aggregation and classification head to allow for the multi-task predictions iii) a training recipe that is successful and produces strong results. See Figure \ref{fig:classification_flow} for a complete flow.

\subsection{Token-line alignment algorithm}

The motivation for our token alignment algorithm is two-fold. Firstly, we observe a flaw with the existing line classification methods \cite{yang2023largelanguagemodelstestfree} for line-level classification, often the \texttt{`\textbackslash n'} (newline) character (and associated token) is used to classify bug lines. This is a problem, however, as tokenizers do not respect the newline boundary. Consistently, across tokenizers and models, tested in our work, the newline character appears in combination with many other tokens (sometimes spanning several newlines) See Figure \ref{fig:new_line_tokens}. Secondly, we want to allow the model to be able to use more than one token of a given line (if needed) to make final predictions if a given line is buggy.

Therefore, in our case we need to associate tokens with specific buggy lines of code and the difficulty lies in that tokens start and end characters often do not neatly correspond to new lines. To solve this issue we develop our own token-line alignment algorithm, which employs a two-pointer like approach. See Algorithm \ref{alg:token-line-alignment} in Appendix \ref{app:token_alignment}.

The algorithm efficiently process tokens and their associated start end ending character indexes and outputs a list (corresponding to lines) of lists of token ids (corresponding to tokens in the given line). See Appendix \ref{app:token_alignment} for more details.

\subsection{Multi-task bug classification head}
Our architecture builds upon a pre-trained LLM backbone, augmented with an additional classification head and optional additional components for bug detection. Importantly, the original backbone LLM can be still used normally during inference with negligible overhead\footnote{The overhead is a single forward pass of the original softmax layer.}.


\subsubsection{LLM backbone}
We employ a pre-trained (code) LLMs (e.g., Qwen-2.5-Coder-7B\footnote{Accessed from here \url{https://huggingface.co/Qwen/Qwen2.5-Coder-7B}.}) as our feature extractor. Given an input of code, the model's tokenizer produces a sequence of tokens $X = [x_1, x_2, \dots, x_n]$, the backbone produces hidden state representations $H = [h_1, h_2, \dots, h_n] \in \mathbb{R}^{n \times d}$, where $d$ is the hidden dimension size. The backbone parameters remain frozen during training to preserve pre-existing code understanding capabilities. We experiment with both just using the additional decoding head as well as adding additional parameter efficient LORA adapter layers (PEFT) \cite{hu2021loralowrankadaptationlarge}.

\subsubsection{Token aggregation}
Our token alignment algorithm gives us the token ids for each line of code. Thus, we aggregate the hidden states of each line of tokens to produce a line-level representation. Given token indices $\mathcal{I}_l$ corresponding to line $l$, we compute:
\begin{equation}
   h_l^{\text{agg}} = \text{Aggregate}(\{h_i : i \in \mathcal{I}_l\}) 
\end{equation}

We experiment with different $\text{Aggregate}(\cdot)$ functions. (1) \textbf{Sum}: the summation of token embeddings, (2) \textbf{Mean}: average of token embeddings, or (3) \textbf{Last-token}: the hidden state of the final token in the line. Our modular architecture allows for other aggregation strategies in future research.

\subsubsection{Decoding (classification) head }

The aggregated line embeddings $h_l^{\text{agg}}$ are passed through a final classification head to predict bug probabilities. The line embeddings runs through an optional feedforward network (FFN) with RELU activation.

\begin{equation}
s_l = \text{FFN}(h_l^{\text{agg}})
\end{equation}

After that we have a final linear projection and classification head for the logits of the new vocab size of 2 (\texttt{NO BUG} and \texttt{BUG}).
\begin{equation}
    p_l = \sigma(W s_l + b)  
\end{equation}
Where $\sigma$ is either \textit{softmax} or \textit{sigmoid}. In our case, $p_l$ is two dimensional and each dimension represents the probability of the two resulting tokens. In our case we mostly do greedy decoding and therefore choose \texttt{argmax} between these tokens.

\subsection{Training recipe}
Line by line prediction generally would happen either on a line by line basis or in batches of lines.
Due to the efficiency of our method we are able to predict all lines for a given file at the same time, as it equates to generating a single token. 

Thus, in our context the notion of \textit{batch-size} means how many files to predict at once (with all the lines in each of the files). It is important to note, that in other works the line number is considered as the batch size \cite{ji2024impactlargelanguagemodels}; compared to such works our batch size is effectively varying from each gradient update to the next. Hence we introduce a weighted loss to account for that.

Concretely, with a batch-size=1 (i.e. one file with many lines) we train our model using a weighted binary cross-entropy loss that processes entire code files simultaneously. For a file with $m$ lines, the loss function is:

\begin{align}
   & \mathcal{L} = - \frac{1}{m} \sum_{l=1}^{m} \nonumber\\ 
   & \left[ w_{\text{pos}} \cdot y_l \log(p_{l,1}) + w_{\text{neg}} \cdot (1 - y_l) \log(p_{l,2}) \right] 
\end{align}

where:
\begin{itemize}
    \item $y_l \in \{0, 1\}$ is the ground truth label for line $l$ (1 if buggy, 0 otherwise)
    \item $p_{l,i}$ is the predicted probability for token 1 and token 2 for line $l$. In the case of softmax $p_{l,1} = 1-p_{l,2}$, in the case of sigmoid these are independent.
    \item $w_{\text{pos}}$ and $w_{\text{neg}}$ are class weights to address label imbalance
\end{itemize}

We choose a single weight based on the validation loss and is treated as a static hyper-parameter. The hyper-parameters can be seen in Appendix \ref{app:hyperparameters}.

\subsection{Summary of method \& hyper-parameters}
In summary our model adds another task or `channel' to the LLM. This allowes the LLM to not only predict the next code token but in parallel also the \texttt{BUG} vs. \texttt{NO BUG} token. This happens via: 1) Embedding each token using an LLM, 2) Aggregating each aligned line token into a single line embedding, 3) using a decoding head to assign probabilities over (in our case two) new tokens, as seen in Figure \ref{fig:classification_flow}. 




%% file: sections/4_experiments.tex
\section{Experimental setup}
\subsection{Datasets}
In order to validate our approach we wanted to test on the well established Java Fault Localization (FL) dataset Defects4J \cite{defects4J}. Additionally, we chose to also use a Python based dataset; to this end we used the PypiBugs dataset \cite{allamanis2021selfsupervisedbugdetectionrepair}. Finally, we also chose to evaluate out-of-distribution (OOD) generalization. For this purpose we created a small python bug dataset based on latest bugs in public projects from Github. We call our dataset BugsEval, which has a total of 12 projects, 121 bugs from these projects and a total of 67 files. We manually annotated all the existing bugs using 2 annotators, see Appendix \ref{app:bugseval} and \ref{app:annotator_instruction}.

In line with previous work \cite{ji2024impactlargelanguagemodels} we split our data into 80\% for training 10\% for validation and 10\% for testing. For full reproducibility we fixed our splits and uploaded them. See details in Appendix \ref{app:dataset_statistics}. 

\subsubsection{Function-level predictions and shorter context}
\label{sec:short_context_exp}
Our work operates on i) the line-level granularity and ii) is able to process entire code files without splitting them up. Line-level granularity is the \textbf{harder task} and also the more useful for downstream tasks. Additionally, being able to process the entire context of a coding file is theoretically more grounded than function-level or chunks, as some bugs depends on the entire file. In fact the entire project would be the best setup in terms of single context, but this exceeds current capabilities of LLMs. Therefore, we believe that our setup is superior in terms of its merits for Automatic Fault Localization research based on LLMs. 

However, to bridge the gap between existing agentic workflows and previous work that operates on chunks of the data we also processed Defects4J similar to LLMAO, by splitting into chunks of roughly 2K tokens. We present these results in Section \ref{sec:short_context_results} as a complete comparison with existing methods, especially LLMAO\footnote{We will release the exact chunked dataset for reproducibility upon acceptance.}.



\subsection{Models}
To include a variety of model sizes we used the strong open source family of Qwen models \cite{qwen2025qwen25technicalreport, yang2025qwen3technicalreport}. The model types we were able to include running on our hardware were: Qwen2.5-Coder-7B, Qwen3-1.7B, Qwen3-4B, Qwen3-8B. Additionally, to show results comparable to previous work we included CodeGen \cite{nijkamp2022codegen}, we included CodeGen 2B, 6B and 16B as it works on our hardware. Notably the CodeGen family of models has only a context length of 2048; therefore our datasets had to be filtered to this length for the CodeGen models. All other experiments ran at a context length of 8000 tokens.

Generally, we train either in BF16 (only possible when we do not do PEFT and only train the decoding head) or 4bit quantization, (4bit is necessary for any PEFT fine-tuning). Inference happens in the BF16 datatype. Additionally all inference happens with default `transformers' \cite{wolf-etal-2020-transformers} settings for all models, see Appendix \ref{app:inference_settings}.

\subsection{Baselines}
In addition to previous work we compare our results to the classical supervised fine-tuning (SFT) paradigm. This baseline is similar to \cite{Wu2023LargeLM}.  The training setup (for comparison) follows the exactly same setup as our method. The main impact is the prompt, which can be found in Appendix \ref{app:sft_prompt}. For the SFT phases we use LORA \cite{hu2021loralowrankadaptationlarge} and 4bit quantization of the models to fit into memory. During inference we run the models in BF16. As software we use the Unsloth and VLLM libraries.

\subsection{Metrics}
Inline with previous work we use the TopK Accuracy. As in \cite{yang2023largelanguagemodelstestfree, ji2024impactlargelanguagemodels} we use K=1,3,5. Specifically, TopK Accuracy is calculated by:

\begin{equation}
\text{Top-}K = \frac{1}{N} \sum_{i=1}^{N} max \left ( \bigcup_{i=1}^{K} \{ \hat{y}_{i,j_k} == {y}_{i,j_k}^* \} \right )
\end{equation}

Where $j_k$ is the line $j$ with the k-highest probability predicted by the model, e.g. for k=1 it is the line with the 1st highest predicted probability of a bug occurring. ${y}_{i,j_k}^*$ is the true answer for file $i$ and line $j_k$ (i.e. either buggy or not buggy); and $\hat{y}_{i,j_k}$ is the prediction (after argmax) for the $i$-th file and $j_k$-th line.

In the case of the SFT model which predicts a list of predicted lines; we do a beam search with beam width of k.

\subsection{Hardware setup}
All our experiments are conducted on 1 x L40S GPUs with 48GB of VRAM running on Ubuntu 22.04. The total GPU hours used was around 384 hours.

%% file: sections/5_results.tex
\section{Results}
We present two main results tables Table \ref{tab:defects4j_results} and Table \ref{tab:pypi_results}. We report Top1, Top3, and Top5 Accuracy in line with previous work. This setting indicates our most realistic setting and the baselines follow the most similar setup. The setting is that: i) the full file is given for prediction, ii) the context length of 8K covers most of the datasets. 

\subsection{Defects4J results}
Looking at Table \ref{tab:defects4j_results} we can see results on Defects4J.
Firstly, we observe that for this comparable setting our approach produces the strongest results on Defect4J, we are able to identify 16.3\% of all bugs with Top1 with our PEFT trained model and 39.5\% with Top5. This is a very significant result as the model has only 1.7B parameters and allows for a context length of 8K on our setup. Inference time is the same as only predicting a single output token (so it is the same to Time to First Token). 

Another interesting observation is that for Defects4J just using a decoding head (without fine-tuning, e.g. DeepFL or PEFT) does not produce strong TOP 1 results. This indicates that the projects are complex and require specialized feature vectors to make successful predictions.

Interestingly, we can also see that the SFT approach does not yield strong results. In fact SFT is among the worst approaches and achieves 0\% with Top1 on the test set. Note the SFT approach achieves a Top1 performance 21.7\% on the training set. Therefore the SFT method `learns' very effectively, however, is unable to go beyond memorization. This is particularly significant as this goes against the common wisdom of "SFT on a specific domain will produce good results".

\begin{table}[t]
\centering
\caption{Comparison of Fault Localization Methods on Defects4J Benchmark. Results are taken from existing work for Ochiai and DeepFL \cite{yang2023largelanguagemodelstestfree}; all other results are single run. QwenCode7B=Qwen2.5, all other Qwen Models are Qwen3.}
\label{tab:defects4j_results}
\begin{tabular}{p{2.9cm}ccc}
\toprule
\textbf{Method}  & \textbf{Top-1} & \textbf{Top-3} & \textbf{Top-5} \\
\midrule
\textbf{Existing Work} &  &  &  \\
\midrule
Ochiai & 4.8\%  & 16.5\% & 25.1\%\\
DeepFL & 14.4\% & \textbf{24.1}\% & 34.2\% \\
\midrule
\textbf{Baseline} &  &  &  \\
\midrule
SFT QwenCode7B & 0.0\% & 7.4 \% & 14.3\% \\
\midrule
\textbf{Our Method} &  &  &  \\
\midrule
MLC QwenCode7B & 9.3\% & 18.6\% & 27.9\% \\
MLC Qwen1.7B & 4.7\% & 14.0\% & 25.6\% \\ 
MLC Qwen4B & 7.0\% & 20.1\% & 30.2\% \\ 
MLC Qwen8B & 7.0\% & 20.1\% & 34.9\% \\
\midrule
\textbf{Ours + PEFT} &  &  &  \\
\midrule
MLC Qwen1.7B & \textbf{16.3}\% & 23.3\% & \textbf{39.5}\% \\
\bottomrule
\end{tabular}
\end{table}

\subsection{PypiBugs results}
Looking at Table \ref{tab:pypi_results} we can see results on PypiBugs\footnote{We note that previous work has not used this dataset for evaluation purposes on the line level, we therefore use only our own baselines.}

Firstly, we observe comparable results to Defects4J indicating that the benchmark has similar difficulty for the models. We also observe that again our PEFT model is the best performing. Interestingly, in this case Qwen2.5-Coder has stronger relative performance vs. Qwen3 than on Defects4J. This indicates that perhaps Qwen2.5-Coder was more exposed to Python than Java. 

Finally, we again observe the poor performance of the classical SFT approach.

\begin{table}[t]
\centering
\caption{Comparison of Fault Localization Methods on Pypi Benchmark. All results are single run. QwenCode7B=Qwen2.5, all other Qwen Models are Qwen3.}
\label{tab:pypi_results}
\begin{tabular}{p{2.8cm}ccc}
\toprule
\textbf{Method}  & \textbf{Top-1} & \textbf{Top-3} & \textbf{Top-5} \\
\midrule
\textbf{Baseline} &  &  &  \\
\midrule
SFT QwenCode7B & 2.0\% & 10.0\% & 14.0\% \\
\midrule
\textbf{Our Method} &  &  &  \\
\midrule
MLC QwenCode7B & 6.3\% & 21.7\% & 28.6\% \\
MLC Qwen1.7B & 6.3\% & 21.7\% & 26.9\% \\
MLC Qwen4B & 9.1\% & 21.7\% & 34.3\% \\ 
MLC Qwen8B & \textbf{10.3}\% & 26.3\% & 36.6\% \\ 
\midrule
\textbf{Ours + PEFT} &  &  &  \\ 
\midrule
MLC Qwen1.7B & 8.6\% & \textbf{27.4}\% & \textbf{37.1}\% \\ 
\bottomrule
\end{tabular}
\end{table}

\subsection{Comparison against function-level and shorter context}
\label{sec:short_context_results}
As discussed in Section \ref{sec:short_context_exp}, we introduce this section to try and bridge the gap between our proposed method and methods that predict on the function level (agentic state-of-the-art methods) and line-level methods that use much smaller context sizes and therefore are also akin to `function' level granularity. 

We accumulate results in Table \ref{tab:defects4J_results_small_context}. The able can be seen as having two distinct set of results. Firstly, we have three agentic systems, which constitute the latest work and strongest results. These scores are based on the `function-level' granularity of predicting bugs.
Secondly, the table then presents LLM based approaches that focus on line-level granularity. LLMAO, being the only LLM based work that has this approach besides ours, has a shorter context (2048 tokens). Therefore we produce two more sets of results using our MLC method using a reduced context also. 

Firstly, we can see that the agentic setup MemFL using GPT-4.1 achieves an impressive 50\% Top1 score and 73.9\% Top5 score on the function-level. LLMAO achieves 46.3\% Top5 with CodeGen 16B. Our results on the other hand indicate a very strong performance and impressive performance. With out best model CodeGen16B we are able to produce a Top5 score of 72\% which is in line with the best agentic method out there. Interestingly, we observe that our Top1 performance for the 6B model is 12.4\%, which is relatively low, while having a high Top5 performance of 42.9\% indicating an inherent ambiguity of the task. We hypothesize that this is due to the chunking of the data and that it can introduce additional ambiguity to the task.

\begin{table}[h]
\centering
\caption{Investigation of TOP-1 and Top-5 results on Defects4J v.1.2.0 for function-level granularity and short-context approaches. AgentFL is taken from \cite{Qin2024AgentFLSL}, it uses GPT-3.5-turbo.  FlexFL is taken from \cite{xu2025flexflflexibleeffectivefault} and uses GPT-4. MemFL is taken from \cite{Yeo2025ImprovingLF}, it uses GPT-4.1-mini. 
CG6B,CG16B = refer to CodeGen 6B and 16B respectively. 
}
\label{tab:defects4J_results_small_context}
\begin{tabular}{lcc}
\toprule
\textbf{Method} & \textbf{Top-1} & \textbf{Top-5} \\
\midrule
\textbf{Function-Level} (<1K) &    &  \\
\midrule
AgentFL (gpt-3.5) & 39.7\% & 47.3\% \\ 
FlexFL (gpt-4) & 49.9\%  & 66.9\% \\
MemFL (gpt-4.1) & 50.9\%  & \textbf{73.9}\% \\
\midrule
\textbf{Short-Context (2K)} &    &  \\
\midrule
LLMAO CG6B & 21.5\% &  40.5\% \\
LLMAO CG16B & 22.3\% &  46.3\% \\
\midrule
\textbf{Ours Short (2K)} &  &  \\
\midrule
MLC CG2B & 14.3\% & 42.9\% \\ 
MLC CG6B & 12.5\% & 42.9\% \\ 
MLC CG16B & 28.6\% & \textbf{71.4}\% \\ 
\bottomrule
\end{tabular}
\end{table}

\subsection{Summary of results}
In this section we saw that our method is significantly better than comparable methods in the full setting of line-level granularity and long-context full files on Defects4J. Additionally, we established new scores on the PypiBugs dataset. 
Interestingly we observed that the classical LLM SFT approach performs terribly for Fault Localization, indicating that alternatives like our multi-task paradigm are needed.
Finally, we also demonstrated superior and comparable performance to previous state-of-the art models in comparable settings, while using much smaller models and orders of magnitude less computation.

%% file: sections/6_analysis.tex
\section{Analysis \& ablations}
\label{sec:analysis}
In this section we further want to study the effectiveness of our method by analyzing generalization of our method as well as the effect of the various hyper-parameters in our setup.

\subsection{Generalization}

We investigate how well our method generalizes. This is particularly important for real-world scenarios where training data may not fully represent the target environment. To this end we evaluate our models by training them on PypiBugs and evaluating on our own BugsEval. The settings remain the same as for the main experiments.

In Table \ref{tab:pypi_to_bugseval} one can see that while Top3 and Top5 performance drop from in-domain scores (see Table \ref{tab:pypi_results}), Top1 performance is actually higher in some cases. This is an interesting observation and indicates that core features of bug localization generalize well to new tasks, while capturing the full breadth of new tasks becomes harder. Again, we also observe that Qwen2.5-Code has a relatively higher performance vs. the Qwen3 model families on this task, again indicating that Qwen2.5 has seen more exposure to python. Finally, we also again observe that PEFT fine-tuning produces even stronger results (Top5 + 5\% points).

\begin{table}[h]
\centering
\caption{Investigating the generalization from Pypi to BugsEval.}
\label{tab:pypi_to_bugseval}
\begin{tabular}{lccc}
\toprule
\textbf{Method} & \textbf{Top-1} & \textbf{Top-3} & \textbf{Top-5} \\
\midrule
\textbf{Our Method} &  &  &  \\
\midrule
MLC QwenCode7B & \textbf{13.8}\% & 21.5\% & 27.7\% \\ 
MLC Qwen1.7B & 10.8\% & 21.5\% & 24.6\% \\
MLC Qwen4B & \textbf{13.8}\% & 18.5\% & 26.2 \% \\ 
MLC Qwen8B &\textbf{13.8}\% & \textbf{23.1}\% & \textbf{29.2}\% \\ 
\midrule
\textbf{Ours + PEFT} &  &  &  \\
\midrule
MLC Qwen1.7B & 9.2\% & 21.5\% & \textbf{29.2}\% \\ 
\bottomrule
\end{tabular}
\end{table}



\subsection{Analysis of key hyper-parameters}


We examine the main hyper-parameters and their effect on the score.\footnote{Note, the main experiments followed the hyper-parameters as seen in Appendix \ref{app:hyperparameters}.} We examined the three most important parameters: 1) Number of Decoding Layers, 2) Aggregation method and 3) Activation function for the Decoding Head. The results can be found in Table \ref{tab:hyperparam_tuning}, \ref{tab:hyperparam_tuning_qwen8b}
 
The observations are quite interesting. Firstly, we noted that increasing the number of layers was not necessarily an improvement of the score, indicating that the features from the LLM are rich enough for effective classification. 
Secondly, we observed that the `last' aggregation method often worked well, which surprised us at first, but makes sense in light of causal language modeling. Furthermore, the mean option was very competitive and sometimes better too, indicating that causal language modeling sometimes does not capture all the pertinent information. Finally, we observed the sigmoid outperformed softmax on most setups. This is very surprising to us; as softmax is usually chosen for next token prediction. While this requires further investigation in future work, we hypothesize that this might be due to an inherent ambiguity in bug classification.

\begin{table}[h]
\centering
\caption{Hyperparameter tuning Top5 results for the Qwen1.7B + PEFT model on Defects4J.}
\label{tab:hyperparam_tuning}
\begin{tabular}{@{}lccc@{}}
\toprule
\textbf{Setting} & \textbf{Max} & \textbf{Mean} & \textbf{Std} \\
\midrule
\textbf{Layers} & & & \\
\midrule
\ 0 & 0.349 & 0.126 & 0.115 \\
\ 1 & \textit{0.395} & \textbf{0.169} & 0.122 \\
\midrule
\textbf{Aggregation} & & & \\
\midrule
\ last & \textit{0.395} & 0.160 & 0.111 \\
\ mean & 0.372 & \textbf{0.196} & 0.127 \\
\ sum & 0.326 & 0.082 & 0.091 \\
\midrule
\textbf{Activation} & & & \\
\midrule
\ sigmoid & \textit{0.395} & \textbf{0.163} & 0.122 \\
\ softmax & 0.349 & 0.131 & 0.116 \\
\bottomrule
\end{tabular}
\end{table}

\begin{table}[h]
\centering
\caption{Hyperparameter tuning Top5 results for Qwen8B on Defects4J.}
\label{tab:hyperparam_tuning_qwen8b}
\begin{tabular}{@{}lccc@{}}
\toprule
\textbf{Setting} & \textbf{Max} & \textbf{Mean} & \textbf{Std} \\
\midrule
\textbf{Layers} & & & \\
\midrule
\ 0 & 0.349 & 0.119 & 0.095 \\
\ 1 & 0.302 & \textbf{0.137} & 0.093 \\
\midrule
\textbf{Aggregation} & & & \\
\midrule
\ last & \textit{0.349} & \textbf{0.164} & 0.097 \\
\ mean & 0.279 & 0.142 & 0.080 \\
\ sum & 0.302 & 0.077 & 0.083 \\
\midrule
\textbf{Activation} & & & \\
\midrule
\ sigmoid & \textit{0.349} & \textbf{0.141} & 0.090 \\
\ softmax & 0.302 & 0.115 & 0.096 \\
\bottomrule
\end{tabular}
\end{table}

\subsection{Summary of analysis}

Overall, we demonstrated that our method is able to generalize to unseen projects. Interestingly, we observed that Top1 generalization was good, while Top5 generalization was weaker, indicating that recognizing bugs generalizes, but catching all / more bugs is hard. 
Additionally, we also investigated hyper-parameters. It was surprising to see that often the `last' token aggregation was the best, while sigmoid performed better than `softmax' indicating perhaps an intrinsic ambiguity about bug classification.

%% file: sections/7_conclusion.tex
\section{Conclusion}
In our work we investigated efficient bug classification methods and presented Multi-task Fault Localization (MLC) that is orders of magnitude more efficient that state-of-the-art agent based system and more granular in their prediction. We demonstrated that for the only other similar approach LLMAO ours is more efficient and produces better predictions and better motivated due to the way actual tokenization works (and the wrong assumptions that being made by existing work). Our results show very competitive performance despite being a very light-touch method. 
We also solve the tokenization alignment problem by proposing a concrete token line alignment algorithm, which can benefit the community more broadly for any line level classification or prediciton tasks for LLMs.



%% file: sections/8_other.tex
\section*{Limitations}
Our paper proposes a multi-channel LLM that is able to both do the next token prediction as well as bug classification. Multi-channel LLMs can form a powerful new paradigm for future research as a way of efficiently capturing multiple tasks with conflicting distributions, while sharing many parameters for general purpose knowledge. A conference paper as this one is unable to capture a full study of such multi-channel LLMs, therefore one limitation of our paper is that we are unable to provide a comprehensive analysis of what multi-channel LLMs can really do and what their effect is. We believe such a study would be better suited over several works in any case.

Another limitation of our work, due to practical compute limitations is that we are unable to study much larger LLMs for this particular multi-channel LLM adaptation. This would require a large compute budget which unfortunately we cannot afford. Overall, we expect the efficiency gains to remain as we require a single forward pass only; additionally stronger LLMs will likely have stronger classification performance as well.

Finally, another limitation we identify in our paper is that function-level granularity (i.e. coarser than our study), which is more common with agentic bug identification approaches, requires future work for the study as it requires a different token alignment algorithm that can efficiently capture function boundaries for different programming languages. Therefore, we believe this is an interesting study, yet a study best suited for future work.

\section*{Ethical considerations}
\textbf{Dataset privacy} we used previously published datasets and public repositories that contain code only and no personally identifiable information.

\section*{Acknowledgments}
To be included in the camera-ready version.





%% file: sections/a_appendix.tex
\section{Model hyper-parameters}
\label{app:hyperparameters}
The below hyper-parameters are the ones we actually chose for all our main experiments in the paper. Note, that a later ablation study revealed certain parameters to be superior in some cases, however, to have a standardized approach we used the same hyper-parameters across models and setttings.

\begin{table*}[h]
\centering
\caption{Hyperparameter search space used for model optimization. The selected values for the final model based are also shown.}
\label{tab:hyperparameters}
\vspace{0.2cm}
\begin{tabular}{ccc}
\toprule
\textbf{Hyperparameter} & \textbf{Values Tested} & \textbf{Selected} \\
\midrule
\texttt{LAYERS} & \{0, 1\} & 1 \\
\texttt{BUG\_WEIGHTINGS} & \{25.0, 50.0, 100.0, 150.0\} & 100.0 \\
\texttt{LEARNING\_RATES} & \{0.0005, 0.0001, 0.00005, 0.00001\} & 0.0001 \\
\texttt{BATCH\_SIZES} & \{1, 2, 4\} & 1 \\
\texttt{AGGREGATORS} & \{\texttt{"mean"}, \texttt{"last"}, \texttt{"sum"}\} & \texttt{"mean"} \\
\texttt{ACTIVATIONS} & \{\texttt{"softmax"}, \texttt{"sigmoid"}\} & \texttt{"sigmoid"} \\
\texttt{PEFT\_RS} & \{32, 64, 128\} & 32 \\
\texttt{PEFT\_ALPHAS} & \{32, 64, 128\} & 32 \\
\bottomrule
\end{tabular}
\end{table*}

The hyper-parameters are described below in more detail.
\begin{itemize}
    \item Layers - indicates the number of layers in the final decoding head.
    \item Bug weighting - indicates the ratio between buggy lines vs. non-buggy lines, specifically the actual weight is computed in the reciprocal.
    \item Learning rates - are the learning rate for AdamW optimizer.
    \item The batch-size - is how many files are used (i.e. all the lines from a single file).
    \item Aggregators - indicates how the line token embeddings are aggregated into a single line embedding.
    \item The Actviations - hyper-parameter indicates which activation the final decoding head has.
    \item PEFT Rs and PEFT alphas indicate Paramter Efficient Finetuning Rank and Alpha for the LORA adapters.
\end{itemize}

\section{Token alignment algorithm}
\label{app:token_alignment}
The below is the full implementation of the token alignment algorithm.

\begin{algorithm}[h!]
\caption{Token-Line Alignment Algorithm}
\label{alg:token-line-alignment}
\begin{algorithmic}[1]
\REQUIRE N lines of code. M total tokens. S total bugs. Token start char index and end char index: $T_{} = [(t_{1,start},t_{1,end}), (t_{2,start},t_{2,end}), $\\$ \dots, (t_{n,start},t_{M,end})]$, line breaks char index $L = [l_1, l_2, \dots, l_N]$ where $l_i$ is char index for line start, bug lines $B \subseteq \{1, 2, \dots, S\}$
\ENSURE List of lists Token indexes corresponding to each line (excluding the newline character) $TL = [[i_{1_1},...,i_{n_1}],\dots,[i_{1_N},...,i_{n_M}]]$ where $i_{n_j}$ corresponds to the token index i $\in \{1, 2, \dots, M\}$ which is included in the j-th row.
\STATE Initialize $y_i \gets 0$ for $i = 1, 2, \dots, n$
\STATE Initialize token pointer $p \gets 1$, line pointer $q \gets 1$
\WHILE{$p \leq n$ and $q \leq m$}
    \STATE $line\_start \gets l_q$
    \STATE $line\_end \gets \begin{cases} 
        l_{q+1} - 1 & \text{if } q < m \\
        n & \text{otherwise}
    \end{cases}$
    \IF{$q \in B$}
        \FOR{$i = line\_start$ to $line\_end$}
            \STATE $y_i \gets 1$ \COMMENT{Mark token as buggy}
        \ENDFOR
    \ENDIF
    \STATE $p \gets line\_end + 1$
    \STATE $q \gets q + 1$
\ENDWHILE
\STATE \textbf{return} $Y$
\end{algorithmic}
\end{algorithm}

\begin{lstlisting}[language=Python, caption={Function to find contiguous token indexes based on character matches}, label={lst:find_token_indexes}]
def find_token_indexes(matches, offset_mapping, return_char_indexes=False):
    """
    finds the CONTIGUOUS matches indexes in offset_mapping.
    
    I.e. token_index collation based on character_index matches in the original string.

    E.g. '\n' was discoverd in a string at char offset 61. this function 
    should return, [[x]] , [[y],[z]] 
        - where x all the token_ids are that contain char 61 token (i.e. only one token likely) 
        - where y is all the token_ids that are before char 61 token
        - where z is all the token_ids that are after char 61 token

    INPUT:
    # matches == (type: List[int]) list of char_offsets for the matched chars 
    # offset_mapping == (type: List[List_2(int)]) list of list(size=2) of ints that represent the char_offsets
    
    OUTPUT:
    token_sequence that does not contain the matches
    token_sequence that does contain the matches

    NOTE:
    If there is not a perfect overlap.
    E.g. 
    - matches = [1,2,3], 
    - offset_mapping = [[0,1],[1,6]]

    The return will be:
    [[1]], [[0]]
    """

    char_indexes_matched = []
    char_indexes_not_matched = []
    ind_char_matched = []
    ind_char_not_matched = []

    matched_token_sequences = []
    individual_matched_token_sequences = []

    not_matched_token_sequences = []
    individual_not_matched_token_sequence = []

    total_sequence_length = len(offset_mapping)
    # print(len(offset_mapping))
    # print(total_sequence_length)
    total_matches = len(matches)

    if total_matches == 0:
        individual_not_matched_token_sequence = list(range(total_sequence_length))
        not_matched_token_sequences.append(individual_not_matched_token_sequence)
        # print("exit 1")
        return matched_token_sequences, not_matched_token_sequences

    matches_idx = 0
    sequence_started = True #of unmatched tokens

    # Kind of like a line-sweep algorithm with double indexing.
    for token_idx, (low, high) in enumerate(offset_mapping):
        # print(f"total_tokens:{total_sequence_length} token_idx:{token_idx} | total_matches: {total_matches} match_idx:{matches_idx}")
        curr_val = matches[matches_idx] #curr_val is the char index of the matched string inside the original string (e.g. newline)
        if curr_val >= low and curr_val < high: #char_index "curr_val" is inside the char range of this token..
            
            # means we matched a "our desired token/string" (eg newline) and are now skipping those tokens.
            if individual_not_matched_token_sequence:
                not_matched_token_sequences.append(individual_not_matched_token_sequence)
                char_indexes_not_matched.append(ind_char_not_matched)
            
            # token id matched
            individual_matched_token_sequences.append(token_idx)
            ind_char_matched.append((low,high))

            # reset
            sequence_started = False
            individual_not_matched_token_sequence = [] #reset for the next unmatched group
            ind_char_not_matched = []

            # now find all the other matches for this token range.
            while curr_val >= low and curr_val < high:
                matches_idx += 1
                if matches_idx >= total_matches: #no more matches.
                    individual_not_matched_token_sequence = list(range(token_idx+1,total_sequence_length))
                    if token_idx+1 < total_sequence_length:
                        ind_char_not_matched = [(offset_mapping[token_idx+1][0], (offset_mapping[total_sequence_length-1][1]))]
                   
                    if individual_not_matched_token_sequence: #could be empty
                        not_matched_token_sequences.append(individual_not_matched_token_sequence)
                        char_indexes_not_matched.append(ind_char_not_matched)

                    matched_token_sequences.append(individual_matched_token_sequences)
                    char_indexes_matched.append(ind_char_matched)
                    # print("exit 2")
                    if return_char_indexes:
                        return return_clean_chars(char_indexes_matched), return_clean_chars(char_indexes_not_matched)
                    else:
                        return matched_token_sequences, not_matched_token_sequences
                    
                curr_val = matches[matches_idx] 

        else: #case where there is no match
            individual_not_matched_token_sequence.append(token_idx)
            ind_char_not_matched.append((low,high))

            if not sequence_started:
                matched_token_sequences.append(individual_matched_token_sequences)
                char_indexes_matched.append(ind_char_matched)

                sequence_started = True
                individual_matched_token_sequences = [] #resetting group of matched tokens.
                ind_char_matched = []
    # at the end we still need to add the last token sequence.
    if individual_not_matched_token_sequence:
        not_matched_token_sequences.append(individual_not_matched_token_sequence)
        char_indexes_not_matched.append(ind_char_not_matched)
    # print("exit 3")
    if return_char_indexes:
        return return_clean_chars(char_indexes_matched), return_clean_chars(char_indexes_not_matched)
    else:
        return matched_token_sequences, not_matched_token_sequences
\end{lstlisting}

\section{SFT prompt}
\label{app:sft_prompt}
For the SFT fine-tuning we use this simple prompt and output format.

\begin{lstlisting}[language=Python, caption={Functions to produce the SFT prompt}, label={lst:sft_prompt_template}]
def prompt_v1(file_source):
    return f"""Please analyze the following code for bugs. Respond only with 'Bug Identified: Yes/No; Lines: <line numbers>' if bugs exist, or 'Bug Identified: No' if no bugs exist.

Code:
```python
{file_source}
```

Now produce the prediction:"""

def output_v1(bug_present, lines=None):
    if bug_present:
        gold_output = f"""Bug Identified: Yes; Lines: {lines}"""
    else:
        gold_output = f"""Bug Identified: No"""
    return gold_output
\end{lstlisting}

\section{Dataset statistics}
\label{app:dataset_statistics}
Previous works have not published their exact data splits, or have included fragmented datasets; our splits will be released on acceptance of the paper.
See Table \ref{tab:dataset_statistics} for exact splits.


\begin{table*}[h!]
    \centering
    \begin{tabular}{cccc}
    \toprule
        Dataset & Train samples & Validation samples & Test samples\\
        \midrule
        Defect4J v1.2 &  1499 & 187 &188\\
        PypiBugs & 288 &19 &78 \\
        BugsEval & -&-&68 \\
        \bottomrule
    \end{tabular}
    \caption{Dataset statistics for all datasets. The number of files is reported.}
    \label{tab:dataset_statistics}
\end{table*}


\section{Inference settings}
\label{app:inference_settings}
Additionally to model hyper-parameters we initialize models as follows using the `transformer' library. The model settings were loaded automatically and the software version was 4.57.2.


\begin{lstlisting}[language=Python, caption={Functions to produce the SFT prompt}, label={lst:sft_prompt_template}]
def load_model_and_tokenizer(
        model_name="Qwen/Qwen2.5-Coder-7B-Instruct", 
        dtype="auto", 
        use_flash_attention=False,
        use_memory_efficient_training=False,
    ):
    """
    Loading HF model (optionally as PEFT)
    """
    print(f"[model loading] Loading: {model_name}")
    tokenizer = AutoTokenizer.from_pretrained(model_name, trust_remote_code=True)

    if use_memory_efficient_training:
        quantization_config = BitsAndBytesConfig(
            load_in_4bit=True,
            bnb_4bit_quant_type="nf4",
            bnb_4bit_compute_dtype=torch.bfloat16,
            bnb_4bit_quant_storage=torch.bfloat16,
        )
    else:
        quantization_config = None

    if use_flash_attention:
        model = AutoModelForCausalLM.from_pretrained(
            model_name,
            dtype=dtype, 
            device_map="auto",
            attn_implementation="flash_attention_2", #trying to reduce OOM and increase speed.
            quantization_config=quantization_config,
            trust_remote_code=True,
        )
    else:
        model = AutoModelForCausalLM.from_pretrained(
            model_name,
            dtype=dtype, 
            device_map="auto",
            quantization_config=quantization_config,
            trust_remote_code=True
        )   
    return model, tokenizer
\end{lstlisting}

\section{Reproducibility}
\label{app:reproducibility}
\begin{itemize}
    \item \textbf{Code Availability:} Our code and models will be available on Publication.
    \item \textbf{Data:} We use publicly available datasets: Defects4J and Pypi Bugs. Our exact splits will be released on publication for full reproducibility.
    \item \textbf{Compute:} Experiments were run on L40S (48GB) with Ubuntu 22.04 (code requirements are released in the code).
    \item \textbf{Hyperparameters:} All hyperparameters are detailed in Table~\ref{tab:hyperparameters}.
\end{itemize}

\section{Licenses}
All datasets and software was used with intended purposes. On release we will release our license as well and clarify usage.
\subsection{Dataset}
\textit{Defects4J} \cite{defects4J} is licensed under open source license ``MIT''.

\textit{PypiBugs}\cite{allamanis2021selfsupervisedbugdetectionrepair} we could not find the license, however the repository and downloads are freely available from Github and Microsoft.


\subsection{Software}
All software is installable from the Pypi repository\footnote{\url{https://pypi.org/}, last accessed May 2026.}

\textit{Unsloth} \cite{unsloth} licensed under open source licenses ``Apache 2.0'' and ``APGL v3''.

\textit{vLLM} \cite{kwon2023efficient} licensed under open source license ``Apache 2.0''.

\textit{transformers} \cite{wolf-etal-2020-transformers} licenced under open source license ``Apache 2.0''.

\textit{Python} (v3.10 and v3.11) licenced under open source license ``Python Software Foundation License''.

\section{Annotator instructions}
\label{app:annotator_instruction}
Two of our direct colleagues, who are domain experts, were instructed to annotate each file with all the lines that contain a bug. No specific instruction sheet or software was provided as the task is `self-explanatory' to domain experts.

The work conducted by the colleagues was done as part of the normal employment.

\section{Impact and risk discussion}
\label{app:impact}

This paper presents work whose goal is to advance the field of machine learning. There are many potential societal consequences of our work, none of which we feel must be specifically highlighted here.

The only exception to the above statement is a minor positive impact that we hope this work might have, which a reduction of energy consumption and thus CO2 consumption through the promotion of better, faster and more efficient machine learning methods.

\section{AI usage disclosure}
\label{app:ai_usage}
During the preparation of this work, the authors used LLMs for language polishing and editing assistance. After using the LLMs, the authors reviewed and edited the content as needed and take full responsibility for the content of the publication.

LLMs were additionally used to find content, but not to create new content or ideas. 

Finally, small coding assistance was used from LLMs, however, most of the is original and written solely by humans.

\section{BugsEval dataset}
\label{app:bugseval}
The dataset will be released upon publication. The projects used were:
fastapi, ollama, tqdm, textual, manim, core, localstack, OpenHands, aider, bugtest-web-app, squash, fullstack-fastapi-template

\section{Future work}
\label{app:future_work}
We believe our work is a stepping stone for a few interesting future research directions. Firstly, it would be interesting to extend our tokenization alignment algorithm to the function level granularity to have a direct comparison against existing (very expensive) agent based methods. Secondly, the multi-channel paradigm opens whole new direction of LLM research where the LLM is trained to predict more than just one next token via multiple decoding heads at the same time. Finally, another future direction is to investigate larger models, which requires a more significant compute budget.

